\documentclass{article}

\usepackage{arxiv}

\usepackage[utf8]{inputenc} 
\usepackage[T1]{fontenc}    
\usepackage{hyperref}       
\usepackage{url}            
\usepackage{booktabs}       
\usepackage{amsfonts}       
\usepackage{nicefrac}       
\usepackage{microtype}      
\usepackage{lipsum}		
\usepackage{graphicx}
\def\BibTeX{{\rm B\kern-.05em{\sc i\kern-.025em b}\kern-.08em
    T\kern-.1667em\lower.7ex\hbox{E}\kern-.125emX}}
\usepackage{doi}
\usepackage{tabularx}
\usepackage{acronym}

\acrodef{iot}[IoT]{Internet of Things}
\acrodef{mno}[MNO]{Mobile Network Operators}
\acrodef{urllc}[uRLLC]{Ultra Reliable and Low Latency Communications}
\acrodef{cpps}[CPPS]{cyber-physical production systems}
\acrodef{qos}[QoS]{Quality of Service}
\acrodef{nm}[NetEm]{network emulation}
\acrodef{nin}[NiN]{Network-in-Network}
\acrodef{ncs}[NCS]{Networked Control System}
\acrodef{embb}[eMBB]{Enhanced Mobile Broadband}
\acrodef{kpis}[KPIs]{Key Performance Indicators}
\acrodef{dsm}[DSM]{Dynamic Spectrum Management}
\acrodef{cnc}[CNC]{Computerized Numerical Control}
\acrodef{fpga}[FPGA]{Field Programmable Gate Array}
\acrodef{cpf}[CPF]{control plane fabric}
\acrodef{dht}[DHT]{distributed hash table}
\acrodef{npn}[NPN]{Non-Public Network}
\acrodef{ai}[AI]{Artificial Intelligence}
\acrodef{rpi}[RPI]{Raspberry Pi}
\acrodef{sm}[SM]{Spectrum Manager}
\acrodef{snc}[SNC]{sub-network controller}
\acrodef{dt}[DT]{Digital Twin}

\title{Dynamic Spectrum Management for 6G Network-in-Network Concepts}

\author{ {Daniel~Lindenschmitt}\\
	Institute for Wireless Communication \\and Navigation\\
	RPTU Kaiserslautern-Landau\\
	\texttt{daniel.lindenschmitt@rptu.de} \\
	\And
	{Paul~Seehofer} \\
	Institute of Telematics\\
	Karlsruhe Institute of Technology\\
	\texttt{paul.seehofer@kit.edu} \\
    \And
	{Marius~Schmitz} \\
	Institute for Manufacturing Technology \\and Production Systems\\
	RPTU Kaiserslautern-Landau\\
	\texttt{marius.schmitz@rptu.de} \\
     \And
	{Jan~Mertes} \\
	Institute for Manufacturing Technology \\and Production Systems\\
	RPTU Kaiserslautern-Landau\\
	\texttt{jan.mertes@rptu.de} \\
 	\And
	{Roland~Bless} \\
	Institute of Telematics\\
	Karlsruhe Institute of Technology\\
	\texttt{roland.bless@kit.edu} \\
    \And
	{Matthias~Klar} \\
	Institute for Manufacturing Technology \\and Production Systems\\
	RPTU Kaiserslautern-Landau\\
	\texttt{matthias.klar@rptu.de} \\
  	\And
	{Martina~Zitterbart} \\
	Institute of Telematics\\
	Karlsruhe Institute of Technology\\
	\texttt{zitterbart@kit.edu} \\
    \And
	{Jan C.~Aurich} \\
	Institute for Manufacturing Technology \\and Production Systems\\
	RPTU Kaiserslautern-Landau\\
	\texttt{jan.aurich@rptu.de} \\
 \And
	{Hans D.~Schotten}\\
	Institute for Wireless Communication \\and Navigation\\
	RPTU Kaiserslautern-Landau\\
	\texttt{schotten@rptu.de} \
}

\date{}

\renewcommand{\shorttitle}{\textit{arXiv} Template}

\begin{document}
\maketitle

\begin{abstract}
Flexible, self-organizing communication networks will be a key feature in the next mobile communication standard. Network-in-Network (NiN) is one important concept in 6G research, introducing  sub-networks tailored to specific application requirements. These sub-networks may be dynamic, i.e., they may appear, disappear, or even move throughout the network. 
Moreover, sub-networks may operate within a shared frequency spectrum, thereby requiring coordination among them. We demonstrate the concept of Dynamic Spectrum Management (DSM) for future 6G networks that dynamically (re-)allocates spectrum according to active sub-networks in the shared spectrum domain. Resilient control plane connectivity between sub-networks and the DSM is provided by the self-organizing routing protocol KIRA, enabling the aforementioned coordination. 
This demonstration presents an integrated solution of the DSM concept, providing increased flexibility to support diverse industrial applications and their individual performance requirements simultaneously within the context of a cyber-physical production system (CPPS).
For the sub-networks, we use specifically designed hardware for wireless real-time communication and couple them with a network emulation. By switching sub-networks on and off, one
can see that the DSM dynamically manages the spectrum allocations for them and that KIRA provides the required connectivity.
\end{abstract}


\section{Introduction}
Highly flexible networks will play an important role in wireless communication networks of the future. They enable them to meet the increasing demand for data transmission with ever more specific requirements \cite{Organic6G}. The introduction of \acp{npn} in the 5G standard is an important basis for this. In addition to usual network operators, private companies can now also operate 5G networks in special frequency bands, e.g., for communication in production facilities. This opens up many new applications, but also new challenges. For future 6G-\acp{npn}, it will therefore be necessary to share limited spectrum with all operators and users to further develop the application scenarios \cite{Scaling}, \cite{lindenschmitt2024nomadic}. In the presented demonstration, \ac{dsm} is introduced as a solution concept for the identified problems. It combines previous work in the field of \acp{nin}~\cite{sublayer} and control plane connectivity for 6G~\cite{kira} into an integrated \ac{dsm} solution. 

\section{Background}
\label{background}
\subsection{Network-in-Networks}
The upcoming 6G standard will increasingly focus on flexibility and simplified operation, which will be supported by self-organizing networks, i.e., operating without manual configuration or intervention. An important step towards this are so-called \ac{nin}, in which several mobile networks (sub-networks) are operated in the same coverage area, but with less transmission power, for example. These sub-networks are independent cellular communication systems that can be adapted to specific requirements. They offer significant advantages over non-cellular standards such as WiFi, especially in terms of robust communication due to higher network determinism \cite{sublayer}.
In the \ac{nin} domain, multiple sub-networks can exchange user data via a shared backbone network. Coordination between the various stakeholders is therefore an inherent necessity. 

In this demo we realize \ac{dsm} using a simple centralized approach featuring a centralized \ac{sm} and decentralized \acp{snc} to meet this increased demand for coordination. The \ac{sm} dynamically accommodates sub-networks with different configurations and \ac{qos} requirements or multiple private mobile networks operated by different providers in shared spectrum scenarios. This sub-networks are connected via a master node to the backbone network. On the master node the deployed \ac{snc} sends the sub-network's requests for specific spectrum to the central \ac{sm}, receives a negotiated spectrum configuration and forwards them to the sub-network, thus forming the basis for the exchange of user data. In this demo we chose a centralized approach for its simplicity. A decentralized approach in which the \acp{snc} discover each other and communicate directly in a peer-to-peer-like fashion to reach consensus on how to distribute the available spectrum
would potentially have better resilience, as it does not have the single point of failure of a central \ac{sm}.

Ensuring trustworthy communication between all parties in the network is crucial, especially when user data is exchanged between different operators \cite{RRSharing}. 6G \ac{nin} will enable application-oriented communication that can adapt dynamically to changing requirements through the \ac{dsm}.

\subsection{Connectivity and Discovery using KIRA}
The coordination needed for the \ac{dsm} requires communication through the shared backbone. 
The resilience of the connectivity enabling this communication is especially important for the operation of the \ac{dsm} (and therefore also the operation of the sub-networks). 
It needs to be available even under network dynamics. 
In this demonstrator we use the routing protocol KIRA~\cite{kira} to provide connectivity among individual sub-networks and all other nodes of the backbone.
KIRA provides scalable, resilient, and self-organizing control plane connectivity as basis for a self-organized operation of the \ac{dsm}. 
Additionally, it also supplies a simple discovery mechanism based on an integrated \ac{dht}. 
This allows nodes to register under well-known names enabling sub-networks to discover the central \ac{sm}.


\section{Application Scenario: Cyber-physical production system}
\label{usecase}
The \ac{dsm} with its central \ac{sm} is suitable for a wide range of applications. As manufacturing systems evolve towards Industry 4.0, networking of \ac{cpps} and the integration of data into \acp{dt} are crucial \cite{digitwin23, 5GEnabler18}. However, the different industrial applications have different requirements for communication networks. For example, closed-loop machine control systems have very high requirements for \ac{urllc} but require low data rates \cite{sublayer}. In contrast, capturing sensor data from a process has lower real-time requirements but requires higher data rates. Different configurations of \acp{nin}, such as \ac{embb} or \ac{urllc}, allow them to be customized specifically to their applications.
\begin{figure}
    \centering
    \includegraphics[width=1.0\linewidth]{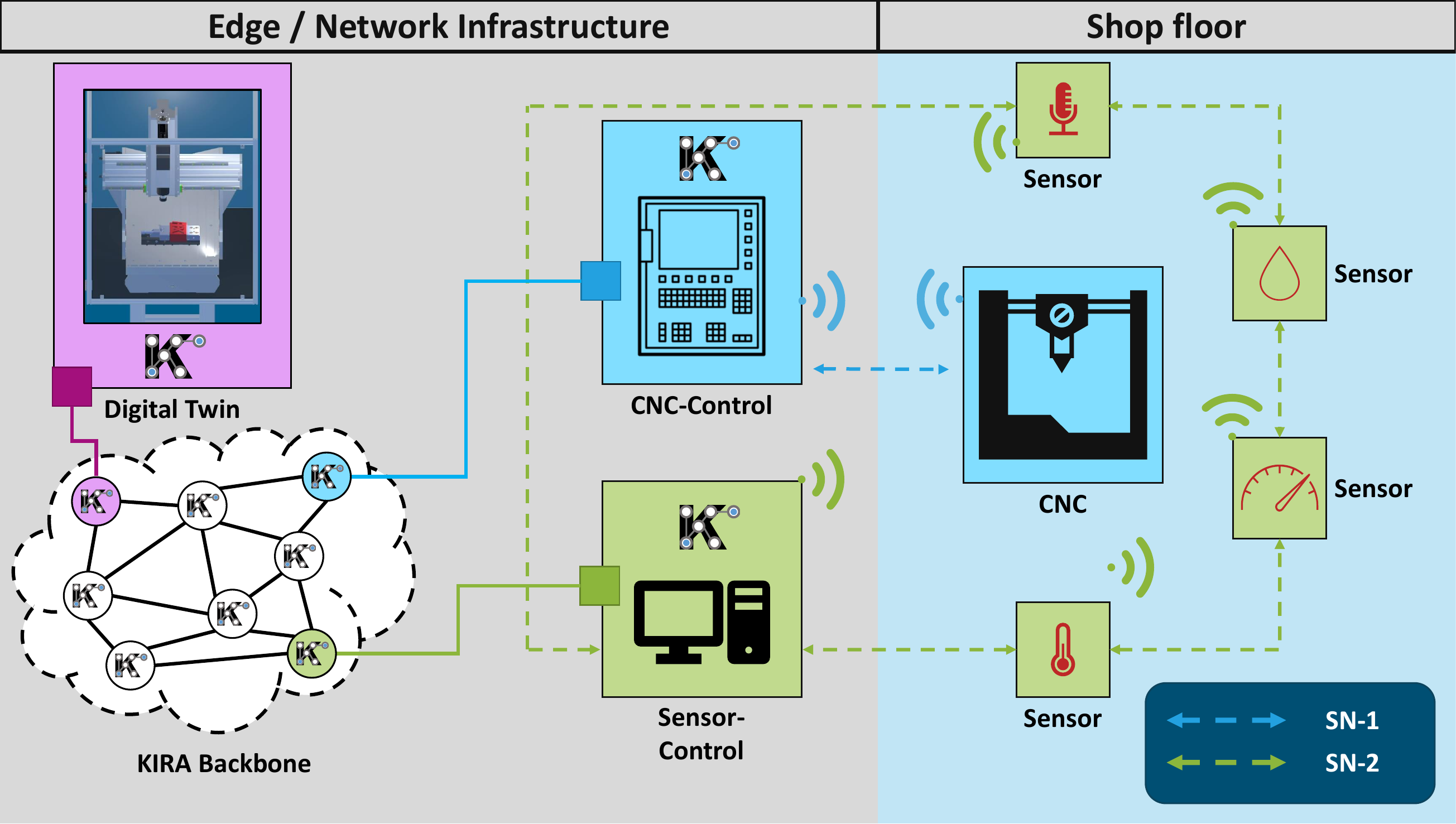}
    \caption{\centering \ac{nin} concept exemplified by a \ac{cnc}-machine with sensors}
    \label{fig:CNC}
\end{figure}

\ac{dsm} enables a new level of reconfigurability and flexibility in \acp{cpps}. Through the central \ac{sm}, sub-networks in highly dynamic \ac{cpps} can be added or removed as needed, ensuring that the available spectrum is always precisely adapted to the applications within a manufacturing system.
Due to its very demanding \ac{kpis}, such as latency and jitter, a \ac{cnc} machine tool featuring a virtualized machine control as part of a \ac{cpps} is considered in this demo. 

As part of a \ac{ncs}, the controller of the \ac{cnc} machine tool is operated on an edge device. The machine considered in this demonstration consists of actuators (servo motors, spindle), sensors (emergency stop, end switches), and a controller and communicate within the bandwidth-limited 5G \ac{npn} with an available spectrum of 100\,MHz between 3.7\,GHz and 3.8\,GHz. The \ac{cnc}s power electronics consisting of sensors and actuators are wirelessly connected to the edge-based controller via specifically designed hardware with a token-based communication system. Simultaneously, the milling process is monitored by various sensors. These sensors ensure that a \ac{dt} has the relevant additional information to monitor the process based on real-time data from the \ac{cnc} controller and to analyze it. In addition to vibration and acoustic sensors, sensors for ambient temperature and humidity were deployed. 

As shown in Figure~\ref{fig:CNC} and Figure~\ref{fig:demo-architecture} the setup consists of two sub-networks, the peer-to-peer sub-network 1 \mbox{(SN-1)} between the \ac{cnc}-controller and the \ac{cnc} machine, and the sub-network 2 (SN-2) for sensor data collection. Both sub-networks share the available spectrum, with two separate networks having individual configurations. As the sensor data in SN-2 require higher data rates, its network uses 60 MHz of the available spectrum, while SN-1 uses the remaining spectrum. 
These sub-networks, and the \ac{dt} host are connected to a shared backbone in which KIRA establishes connectivity enabling communication with the central \ac{sm}.

\section{Demo Setup}
\label{demo}

\begin{figure}[b]
    \centering
    \includegraphics[width=1.0\linewidth]{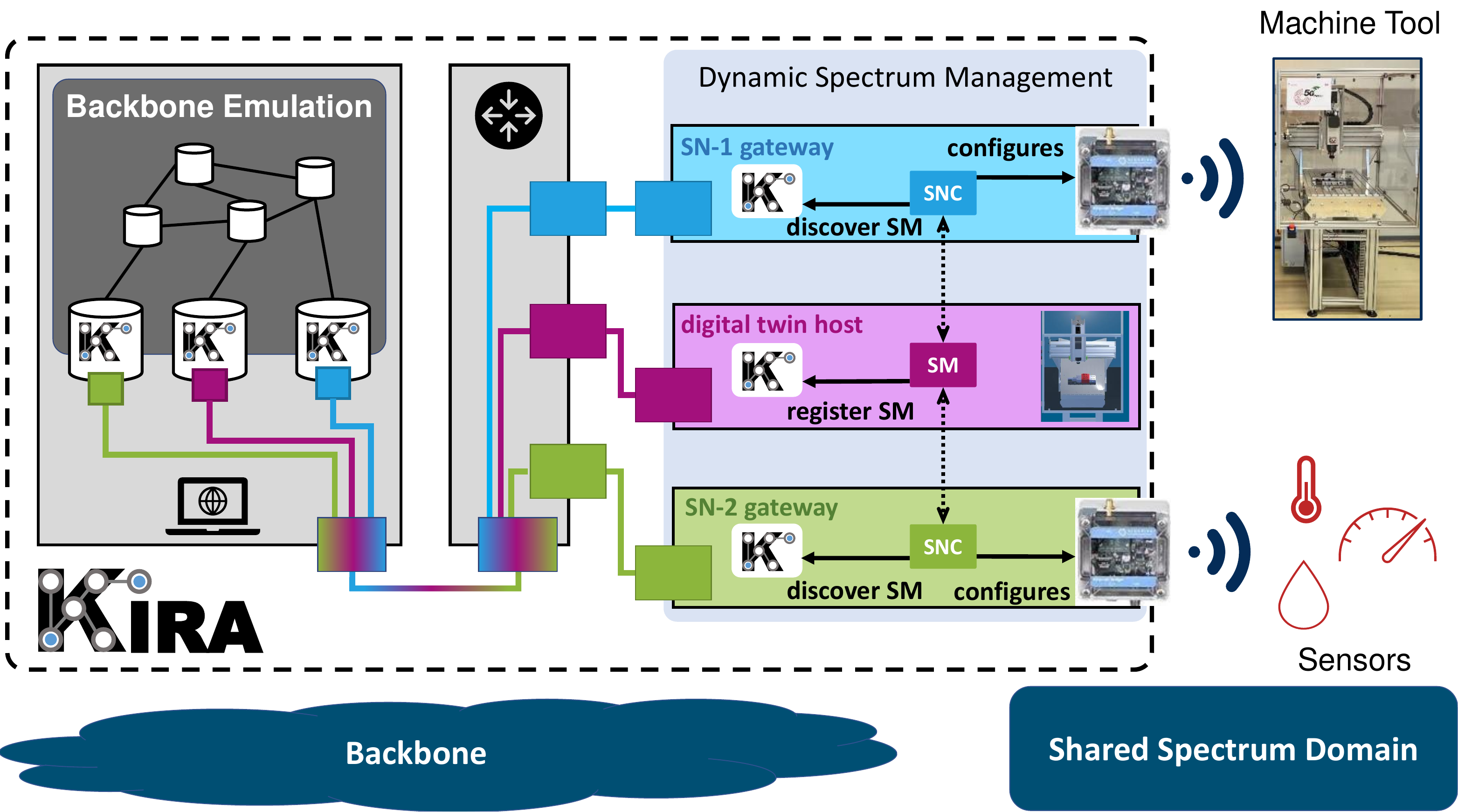}
    \caption{Demo Architecture (SM = spectrum manager)}
    \label{fig:demo-architecture}
\end{figure}

The demonstrator consists of different parts (see Figure \ref{fig:demo-architecture} and \ref{fig:demo-setup}): an emulated backbone network, which can also be implemented as an real hardware deployment, as well as two physical \acp{nin} and an additional physical host running a \ac{dt} of the \ac{cnc} environment.

\subsection{Backbone Network}
The backbone network consists of a set of routers that is emulated using containernet~\cite{containernet} on a separate host. Three containers of the emulated network are connected to one of the physical parts (sub-networks and \ac{dt} node), through virtual links separated through VLANs using a VLAN-capable switch. Within this backbone, all nodes (physical and emulated) run a KIRA routing daemon that establishes IPv6 connectivity between all nodes. The demonstrator features a visualization of the backbone showing how KIRA establishes connectivity and how its \ac{dht} is used to discover the \ac{sm}.

\subsection{Network-in-Networks}
The \acp{nin} are implemented using \acp{rpi}, each \ac{rpi} is acting as a master node for the sub-networks behind it. A \ac{snc} on the \acp{rpi} is responsible for communicating with the \ac{sm} via KIRA, it registers the spectrum requirement, and receives a corresponding response from the \ac{sm} following an internal negotiation process. The actual sub-network, which in this demo consist of specially designed hardware (see Figure \ref{fig:demo-setup}) for data exchange based on a token-based transmission protocol in the frequency range from 3.7\,GHz to 3.8\,GHz, are connected to the \acp{rpi}. The sub-network SN-1 is responsible for the communication between \ac{fpga} and \ac{cnc} machine, and SN-2 for the data exchange of the individual sensors.

\subsection{Self-Organizing Dynamic Spectrum Management}
The \ac{dsm} is realized as a simple client-server architecture: a spectrum manager for simplicity also running on the \ac{dt} node stores its IPv6 address (which was randomly generated by KIRA) in KIRA's \ac{dht} under the key ``dynamic-spectrum-manager'' and waits for connections. The \ac{snc} on the \ac{rpi} performs the respective \ac{dht} fetch and connects to the spectrum manager. The spectrum manager then allocates spectrum according to the currently connected \acp{nin} and sends appropriate commands back to the \acp{snc}, which \mbox{(re-)}\,configure the sub-networks. This way the sub-networks do not need any special prior configuration, e.g., a particular address of the spectrum manager.
As a result, if one of the two networks fails, the remaining network is reconfigured to use the spectrum of the failed network. As shown in Figure \ref{fig:demo-setup}, the distribution of the spectrum is visualized on a dashboard in our demonstration. The \ac{dsm} detects this failure and initiates the reconfiguration of the remaining network.
The following table \ref{tab:spectrum} illustrates the logic of the network of our setup.

\begin{table}[h]
\centering
\caption{DSM Logic of the demonstration}
    \begin{tabularx}{\linewidth}{lcc}
        \hline        
         &   \textbf{CNC Spectrum} & \textbf{Sensor Spectrum}  \\
        \hline
        Both running &  40\,MHZ & 60\,MHz \\
        
        Sensor Network unavailable & 100\,MHz & 0\,MHz \\ 
        
        CNC Network unavailable &  0\,MHz & 100\,MHz \\
        
        Both unavailable &  0\,MHz & 0\,MHz \\
        
        \hline
    \end{tabularx}

    \label{tab:spectrum}
\end{table}

\section{Demo Walkthrough}
\label{walk}
\begin{figure}
    \centering
    \includegraphics[width=1.0\linewidth]{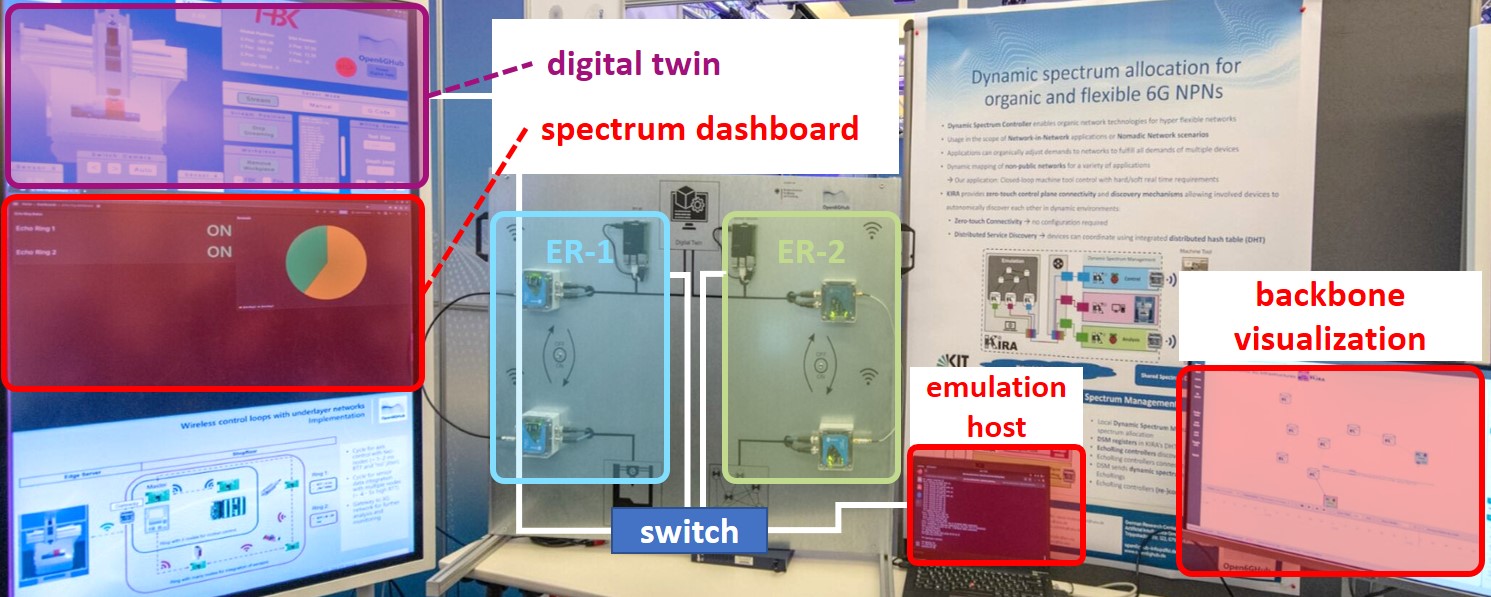}
    \caption{Demo Setup}
    \label{fig:demo-setup}
\end{figure}

The demonstration shows the integration of the \ac{dsm} into a real \ac{nin} setup and the reaction of the system to changes within the network structure (see Figure \ref{fig:demo-setup}). As already explained in Section \ref{demo}, dynamic adaptations to network requirements, e.g. in the production environment, can be realized in this way. The demonstration is divided into three different components:
\begin{itemize}
    \item \ac{nin}: Visitors can interactively switch on/off individual sub-networks and observe resulting adjustments by the \ac{dsm}. The \ac{sm} initiates the reconfiguration of the networks and assign new frequencies, executed by the \ac{snc}.
    \item \ac{dt}: Operational capability and performance of \acp{nin} will be demonstrated as a \ac{dt} using the industry-oriented application of a \ac{cnc} machine with sensor data evaluation.
    \item Visualization: Control and configuration in the shared backbone are displayed both for the \ac{dsm} as a dashboard of the current configuration and for KIRA's connectivity.
\end{itemize}

\section{Conclusion and Future Work}
\label{concl}
In this paper a novel self-organizing \acl{dsm} for 6G networks was introduced and its application in a \ac{nin} setup is demonstrated. The proposed \ac{dsm} efficiently allocates spectrum to fulfill various application requirements in Industry 4.0 and ensures robust communication. The demonstration with a \ac{cnc} machine tool highlights the adaptability and reliability of this approach. Future work will focus on the development of decentralized \ac{dsm} solutions and the integration of more sophisticated dynamic spectrum management techniques to support nomadic networks.
\section*{Acknowledgment}
The authors acknowledge the financial support by the German \textit{Federal Ministry for Education and Research (BMBF)} within the project Open6GHub \{16KISK004\ \& 16KISK010\}.

\bibliographystyle{ieeetr} 
{%
\bibliography{references}
}%
\end{document}